\documentclass[12pt]{article}
\input epsf.tex

\newcommand{\be}{\begin{equation}}
\newcommand{\ee}{\end{equation}}
\newcommand{\bea}{\begin{eqnarray}}
\newcommand{\eea}{\end{eqnarray}}

\newcommand{\gs}{\ensuremath{g_s}} 
\newcommand{\ls}{\ensuremath{l_s}} 
\newcommand{\lP}{\ensuremath{l_P}} 

\def\p{\partial}

\newcommand{\tr}{\mathop{\rm Tr}}
\def\expec#1{\langle #1 \rangle}

\newcommand{\cO}{{\mathcal{O}}}
\newcommand{\cN}{{\mathcal{N}}}

\newcommand{\bT}{\ensuremath{\bar{T}}}

\newcommand{\Tau}{\ensuremath{\mathcal{T}}}

\newcommand{\Dn}{\ensuremath{\overline{\mbox{D9}}}}
\newcommand{\DD}{\ensuremath{\mbox{D-}\overline{\mbox{D}}}}
\newcommand{\bN}{\ensuremath{\bar{N}}} 
\newcommand{\Et}{\ensuremath{M_{\mathrm{FT}}}}  
\newcommand{\Eg}{\ensuremath{E}}  
\newcommand{\Sm}{\ensuremath{S_{\mathrm{FT}}}}
\newcommand{\Ss}{\ensuremath{S_{\mathrm{SG}}}}
\newcommand{\Ms}{\ensuremath{M_{\mathrm{SG}}}}
\newcommand{\Th}{\ensuremath{T_{\mathrm{FT}}}}

\begin{document}

\begin{titlepage}

\begin{flushright}
UUITP-05/01\\
USITP-01-06\\ 
hep-th/0106201
\end{flushright}

\vspace{1cm}

\begin{center}
{\Large\bf Brane-Antibrane Systems at  Finite Temperature\\ 

\smallskip

and the Entropy of Black Branes \\}

\end{center}
\vspace{3mm}

\begin{center}

{\large Ulf H.\ Danielsson,$^{\scriptstyle 1}$
\renewcommand{\thefootnote}{\fnsymbol{footnote}}
Alberto G\"uijosa,$^{\scriptstyle 2}$\footnote[1]{Address 
after Sept.~1: Depto. de F\'{\i}sica de Altas Energ\'{\i}as,
Instituto de Ciencias Nucleares,
Universidad Nacional Aut\'onoma de M\'exico,
Apartado Postal 70-543, 04510, 
M\'exico D.F., M\'exico.}
and
Mart\'\i n Kruczenski$^{\scriptstyle 1}$\footnote[4]{Address 
after Sept.~15: 
Dept. of Physics, Univ. of Toronto, 60 St. George st.,
Toronto ON, M5S 1A7 Canada. 
Also at Perimeter Institute, Waterloo, Canada.}} \\
\renewcommand{\thefootnote}{\arabic{footnote}}

\vspace{5mm}

$^{1}$ Institutionen f\"or Teoretisk Fysik, Box 803, SE-751 08
Uppsala, Sweden

\vspace{3mm}

$^{2}$ Institute of Theoretical Physics, Box 6730, SE-113 85
Stockholm, Sweden

\vspace{5mm}

{\tt
ulf@teorfys.uu.se, alberto@physto.se, \\
martin.kruczenski@teorfys.uu.se \\
}

\end{center}

\vspace{5mm}

\begin{center}
{\large \bf Abstract}
\end{center}
\noindent

We consider 
D-brane--anti-D-brane systems at $T>0$.  Starting at the closed string 
vacuum, we argue that a finite temperature leads to the reappearance 
of open string degrees of freedom.  We also show that, at a 
sufficiently large temperature, the open string vacuum becomes stable.  
Building upon this observation and previous work by Horowitz, 
Maldacena and Strominger, we formulate a microscopic brane-antibrane 
model for the non-extremal black three-brane in ten dimensions (as 
well as for the black two- and five-branes in eleven dimensions).  
Under reasonable assumptions, and using known results from the AdS/CFT 
correspondence, the microscopic entropy agrees with the supergravity 
result up to a factor of $2^{\frac{p}{p+1}}$, 
with $p$ the dimension of 
the brane.  The negative specific heat and pressure of the black brane 
have a simple interpretation in terms of brane-antibrane
annihilation.  We also find in the model states 
resembling black holes and other lower-dimensional 
black branes.

\vfill
\begin{flushleft}
June 2001
\end{flushleft}
\end{titlepage}
\newpage

\section{Introduction}

Polchinski's identification of D-branes \cite{polchrr}
as the microscopic counterparts 
of the RR-charged black brane solutions of type II
supergravities \cite{hs}
opened the possibility to 
explain black brane entropy 
in terms of the open strings living on
the D-branes.
This possibility was realized by Strominger and Vafa~\cite{sv}, 
who were able to reproduce the exact entropy of certain extremal 
black holes through a microscopic state counting. 
The counting was soon extended to some  
near-extremal cases \cite{nearext}, and this was
followed by a tremendous surge 
of activity \cite{entropy}, which led to the discovery of
the AdS/CFT correspondence \cite{malda} as a remarkable by-product,
and continues even today (see, e.g., the interesting 
recent works \cite{juan,jmc,bhrecent}).  

Most of the examples where
a successful microscopic description has been found
involve extremal and near-extremal black branes. 
These systems have positive specific heat, and their
entropy can be accounted for using a gas of finite temperature 
living on the D-branes. 
Black branes which are far from extremality, like the ordinary
Schwarzschild black hole, are in this respect much more challenging.
Consider for example the case of a neutral
black three-brane of mass $M$ wrapped on a torus of volume $V$.
Its entropy is given by \cite{entropy}
\be  \label{d3ent}
S = 2^{\frac{9}{4}} 5^{-\frac{5}{4}} \pi^{\frac{1}{4}}  
    \sqrt{\kappa} M^{\frac{5}{4}}V^{-\frac{1}{4}}~.
\ee
If one tries to interpret this as the entropy of a gas of particles with
energy $E=M$ some well-known problems arise. First, since the 
power of $E$ is larger than one, 
a simple thermodynamical calculation shows that 
the specific heat is negative. Second,
by extensivity the exponent of $V$ is forced to be negative, 
which implies that 
the pressure $p = T\left(\partial S/\partial V\right)_M$ is negative. 
Finally, when the black brane evaporates
completely through Hawking radiation, 
all of its  excitations
disappear, meaning that the
field theory which describes it microscopically 
should have a vacuum with no
degrees of freedom other than (possibly) closed strings. 

The last point suggests that the relevant
field theory should perhaps display
confinement: the black brane could then be associated with an
unconfined 
vacuum, and the endpoint of the evaporation process would
be identified with
the confined vacuum, where flux tubes (closed strings) are the only 
allowed excitations. 
This could also explain the negative pressure: because of
the energy difference between the confined and unconfined vacua,
the system would gain energy by reducing its volume. 

The starting point for the present work is the observation that
these same properties are in fact possessed by 
D-brane--anti-D-brane systems. These systems
were first analyzed in \cite{bs}, 
and have been much studied of late \cite{nonbps}. 
The fact that the corresponding theory has a vacuum with only
closed string excitations was conjectured by Sen \cite{sennonbps}.
The exact mechanism through which open string degrees of freedom 
disappear is still in debate; but a particularly promising proposal 
maintains precisely
that the theory at the vacuum in question is confined \cite{confinedu1},
and that closed strings are to be understood as tubes of the confined 
flux \cite{confinedstrings,ncstrings,senstrings,kls}.

In this paper we will argue that non-dilatonic black branes may be understood 
microscopically as collections of branes and antibranes. 
It had been noted already in \cite{hms} that the Bekenstein-Hawking entropy 
of the D1-D5 system arbitrarily far from extremality can be rewritten in a 
form suggestive of a microscopic
model involving branes and antibranes 
(an observation which was extended to 
other cases in \cite{antientropy}).
However, the understanding of D-$\overline{\mbox{D}}$ systems was at 
that time insufficient to attempt a direct formulation of such a 
model--- in particular, it would have been unclear why the branes and 
antibranes do not annihilate. As we will review in the next section,
the situation in this regard has 
improved considerably in recent years. The significance of the tachyonic 
instability of the D-$\overline{\mbox{D}}$ system is now understood, 
and the exact tachyon potential has been determined 
\cite{mzsuper,kmmsuper}. Furthermore, it has been observed
that on the worldvolume of a brane-antibrane system one can find
solitons describing not only fundamental strings but also 
D-branes of lower dimensionality \cite{sennonbps,dsol,ncstrings}. 
These results support the hope
that the open string field theory associated with a space-filling 
brane-antibrane pair could give
a non-perturbative definition of the full string theory. 
If this is correct, then it should certainly
be possible to compute within this framework
the entropy of black branes.

In the following we will examine these ideas in more detail.
In particular, we will
show that, under reasonable assumptions about the degrees of 
freedom (including their reduction by a factor of $3/4$ due to 
strong-coupling \cite{gkp,threequarters}), 
a model with branes, antibranes, and corresponding 
gases of open strings, can precisely
reproduce the black brane entropy (\ref{d3ent}), 
up to a puzzling factor of $2^{\frac{3}{4}}$ 
which seems difficult to understand. 
We have already observed above that a D-$\overline{\mbox{D}}$
system has the correct properties to explain the disappearance 
of the black brane's degrees of freedom
after its evaporation, as well as its negative pressure.
As we will see, this system can also account for the
negative specific heat.  
The explanation is simply that, when energy is taken from the gas living on
the branes, it is entropically favorable to annihilate some of the 
branes, which effectively increases the temperature of the gas. 
In other words, for a fixed
number of branes the specific heat is of course positive, 
but the possibility of
getting energy from brane annihilation 
(or tachyon condensation in Sen's language) 
makes the specific heat of the whole system negative.

The paper is structured as follows. 
We begin in Section \ref{2sec} 
by discussing generic features of the brane-antibrane system at finite
temperature. We will argue that, starting at the closed string vacuum,
a finite temperature leads to the (re)appearance of open string degrees 
of freedom, a phenomenon which might have implications for the 
nature of the Hagedorn transition. 
We will also show that,
at a high enough temperature, the `tachyon' 
field at the open string vacuum is no longer tachyonic, 
and so the D-$\overline{\mbox{D}}$ system 
is stable. The analysis in later sections relies 
only on this last point, 
so readers interested mostly in black brane entropy 
considerations might wish to skip directly to Sec.~\ref{stablesec},
where the point is made.
In Section \ref{3sec} we will 
formulate a simple D-$\overline{\mbox{D}}$ model
for the black three-brane in ten dimensions
(as well as the black two- and five-branes in eleven 
dimensions), and show that it 
correctly accounts for various properties
of the supergravity solution.    
We discuss in detail the case of a Schwarzschild (neutral)
black brane, and note that for the charged case
supergravity curiously seems to 
suggest that the gases living on the
branes and antibranes have the same energies,
and therefore different temperatures.
Section \ref{4sec} demonstrates that our microscopic model
can reproduce certain features of the Gregory-Laflamme instability 
\cite{GL}.
Our conclusions are given in Section \ref{5sec}.

Other attempts to give a microscopic description of Schwarzschild black holes 
have been made previously, most notably via the string/black hole 
correspondence \cite{stringbh,hp}, and in the Matrix theory 
\cite{bfss} context \cite{bfks,matrixbh,dmrr,boostbh}.

\section{Brane-Antibrane Systems at Finite Temperature}
\label{2sec}

The knowledge emerging from the vigorous investigation of unstable branes 
in recent years \cite{nonbps} has 
enhanced our understanding of the structure of string theory.
This activity was initiated by
Sen \cite{sennonbps}, who in particular conjectured 
that D-brane systems with
a perturbative tachyonic excitation disappear completely via
condensation of the associated
scalar field (the `tachyon'). 
At the end of this process
the theory is defined in a new vacuum, whose excitations are 
described by closed strings alone.

The verification of this proposal requires tools capable of describing 
strings off-shell, and this has led to a resurgence of string field 
theories. Considerable progress has been made along three 
distinct but intersecting routes: 
cubic open string field theory \cite{csft} 
(see in particular the works \cite{tachcsft,dsol,cubicmass}),
noncommutative theories \cite{ncsol,ncstrings} 
(for a review, see \cite{harvey}), 
and background independent or boundary string field theory 
(BSFT)\footnote{This is essentially a refinement of the  
old sigma-model approach. See \cite{sigma} and references
therein.}  
\cite{wbsft,shatbsft}.
Each of these methods has illuminated different 
aspects of the process of tachyon condensation; together they have 
provided an impressive quantitative verification of Sen's proposal. 
These results can be viewed as an indication that
string field theories truly constitute a non-perturbative formulation 
of string physics, and are thus a valuable complement 
to the formulations provided to us by 
Matrix theory \cite{bfss} and Maldacena's duality \cite{malda}.

By means of BSFT, in particular, it has been possible to 
determine the tachyon potential exactly\footnote{Remarkably, 
it had been anticipated in \cite{mz,mzsuper}, where it was shown to 
possess all the desired features.} 
\cite{gs,kmm,kmmsuper}. 
This potential has the property that its second derivative at 
the minimum (i.e., the mass of the tachyon,) is infinite, which 
is in agreement with our expectation that, at the end of the
condensation process,
all open string degrees of freedom are absent--- the D-branes have 
disappeared. 
In this section we will argue, however, that
the description of the closed string vacuum as a tachyon condensate 
has an interesting consequence: if we turn on a finite temperature, 
the open string degrees of freedom will reappear. 
Our main interest is the case of a brane-antibrane system in 
superstring theory, but the essential 
features are
the same for unstable branes in bosonic or supersymmetric 
string theories.

\subsection{A single brane-antibrane pair}
\label{singlesec}

For concreteness, we will carry out the analysis for 
a single D9-\Dn\ pair in type IIB string theory. 
As always, the
excitations of this D-brane system are described by open strings
whose  endpoints are anchored on the branes. 
The 9-9 and $\overline{9}$-$\overline{9}$ strings
give rise to the usual (GSO-even)
gauge fields and gauginos $\{A_{\mu},\Lambda_{\alpha}\}$, 
$\{\overline{A}_{\mu},\overline{\Lambda}_{\alpha}\}$
on the brane and the antibrane. 
The 9-$\overline{9}$ and $\overline{9}$-9 strings, on the other 
hand, yield GSO-odd states: 
a complex tachyon $\phi$ from the NS sector, 
and massless fermions $\Psi_{\alpha},\overline{\Psi}_{\alpha}$ 
from the R sector.
Since it originates from strings running between the brane and the 
antibrane, the tachyon is charged under the relative $U(1)$
(i.e., $A^{-}_{\mu}\equiv A_{\mu}-\overline{A}_{\mu}$),
but is neutral under the overall $U(1)$
($A^{+}_{\mu}\equiv A_{\mu}+\overline{A}_{\mu}$).

\begin{figure}[thb]
\centerline{\epsfxsize=7cm\epsfbox{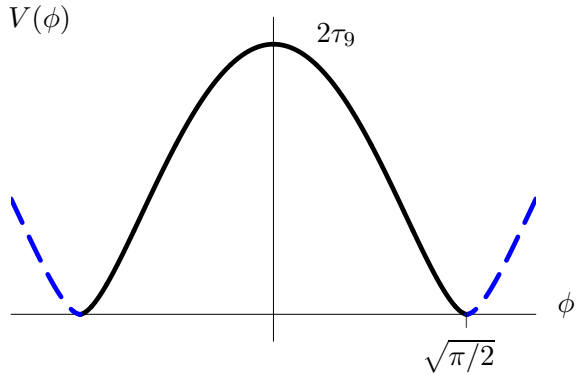}}
 \begin{picture}(0,0)
  \put(41,47){\small $V(\phi)$}
  \put(82,45){\small $2\tau_{9}$}
  \put(114,9){\small $\phi$}
  \put(101.5,7.5){\small $\scriptscriptstyle |$}
  \put(96,2){\small $\sqrt{\pi/2}$}
 \end{picture}
 \vspace*{0.1cm}
\caption{\small Tachyon potential for the brane-antibrane system. The 
physical range for the field 
is $|\phi|<\sqrt{\pi/2}$ ($0\leq |t|<\infty$); the dotted lines outside this 
range have been drawn to aid the eye.}
\end{figure}

The exact
potential for the tachyon 
takes the form \cite{mzsuper,kmmsuper,ddbar}
\be \label{v}
V(\phi)=2\tau_{9}\exp[-2|t(\phi)|^{2}]~,
\ee
where $\tau_{9}$ is the D9-brane tension, and $t$ is the  
background tachyon field
appearing in the worldsheet action, related to
$\phi$ (the field with a standard kinetic term in the BSFT action)
through the error function,
\be \label{tphi}
|\phi|=\sqrt{\pi\over 2}\mbox{Erf}(|t|)~.
\ee
The potential (\ref{v}) is of the `Mexican hat' type (see Fig.~1):
it has a maximum at $\phi=0$ (the open string vacuum) 
and a minimum at $|\phi|=\sqrt{\pi/2}$ (the closed string vacuum).  
Close to the minimum it takes the form
\be \label{vapp}
V(\phi)=-4 \tau_{9}(\sqrt{\pi\over 2}-|\phi|)^{2}
       \ln(\sqrt{\pi\over 2}-|\phi|)+\ldots
\ee

As noted before, 
the fact that the tachyon has an infinite mass at the closed string 
vacuum ($V''(\phi=0)=+\infty$)  is the concrete expression of its
removal from the spectrum of the theory.
This is in fact expected for 
all open string degrees of freedom, although the precise manner in which 
each mode disappears is not entirely clear. 
This is especially true for the gauge field $A^{+}_{\mu}$:
calculations in cubic string field theory suggest that it acquires a 
mass \cite{cubicmass}, whereas ordinary field theory intuition and 
BSFT suggest otherwise \cite{u1problem,kmm} (see also 
\cite{senaction,senstrings,gshiggs,mzgauge}). 
A promising mechanism for its ultimate
removal from the theory is the proposal that the 
associated $U(1)$ becomes confined \cite{confinedu1}.
This picture is attractive because, as a bonus,
it allows one to understand
closed strings at the $\phi=0$ vacuum as
collective excitations of the open string field:
they are interpreted as
electric flux tubes\footnote{The idea that closed strings are electric 
flux tubes has been discussed independently of the confinement 
scenario in \cite{ncstrings,senstrings}.}
of the confined $U(1)$ \cite{confinedstrings,kls}.

If we now consider this same system at a finite temperature, standard 
thermal field theory reasoning tells us what to expect: 
a small temperature should lead to an effective potential 
in which the location of the minimum has shifted away from 
$|\phi|=\sqrt{\pi/2}$. 
The physical reason 
for this is that moving towards $\phi=0$ can be thermodynamically 
favorable: it costs energy, but
it also reduces the mass of the tachyon and 
therefore increases the entropy of the tachyon gas.
The optimal configuration is the one that minimizes the free energy of 
the system, and this will vary with the 
temperature.
For a large enough 
temperature the minimum could conceivably shift 
all the way to $\phi=0$, 
in which case the open string vacuum would be 
stable.

When attempting to analyze this  effect explicitly for the 
potential (\ref{v}), one faces a number of difficulties. 
First, the field theory we are studying is non-renormalizable,
not only because (\ref{v}) describes an interacting 
scalar field in 9$+$1 dimensions,  
but also because the complete action for the tachyon includes
higher-derivative corrections 
(see \cite{kmm,der} and refs. therein). 
A related point is that the field 
theory (\ref{v}) appears to be strongly coupled near the closed 
string vacuum, whereas string theory intuition would suggest that the 
coupling is $e^{-2|t|^{2}}\gs N$, 
which would mean that the system is weakly coupled
for $|\phi|\to \sqrt{\pi/2}$ ($|t|\to\infty$). 
This puzzle was noted in \cite{kmm}
and has been further discussed in \cite{kls,ckl}. 
The way to address
these problems 
would be to work in the full string theory,
including the effect of all the higher open string modes
by performing a one-loop calculation in 
the BSFT language. For the zero-temperature case
this has been attempted in various ways
in \cite{annulus,ckl}; see especially \cite{ckl} 
for a discussion of the issues involved.\footnote{The 
$T> 0$ case is 
examined in \cite{huang}, which appeared while this paper was being 
written.}  

The need to resort to the full string theory
highlights another qualification of our analysis: 
we will in what follows focus attention on the tachyon field alone,
ignoring the contribution
of all other open string modes to the free energy of the system.
The rough picture we have in mind here is that as one approaches the 
closed string vacuum, the entire open string mass 
spectrum is shifted uniformly.\footnote{It is amusing to note that 
this is what a naive application of BSFT on the \emph{cylinder}
would predict: a constant tachyon background $t=a$ 
in the worldsheet action simply multiplies the 
partition function by a factor of 
$e^{-a^{2} l}$ (where $l$ is the length of the boundary), 
which can be interpreted as a uniform mass-squared shift 
for all the modes. {}Given the relation (\ref{tphi})
between $t$ and $\phi$,
this shift  in fact agrees with 
the tachyon mass shift inferred from $V''(\phi)$, 
$M^{2}+1/2=t^{2}$.}
If the tachyon has the lowest mass,
its contribution will be dominant as long as the temperature is 
low compared to the Hagedorn temperature.
Summarizing, we hope that,
despite its shortcomings,
the minimalistic calculation that follows
at least captures
the essential qualitative features of the problem.

The free energy 
for the brane-antibrane $+$ tachyon condensate $+$ gas 
system at a finite temperature $T=\beta^{-1}$  is
\be \label{f}
F(\phi,\beta)=2\tau_{9}e^{-2|t|^{2}}
+{\Omega_{8}\over(2\pi)^{9}}\beta^{-10}\int_{0}^{\infty}dx\,x^{8}
\ln\left[1-e^{-\sqrt{x^{2}+\beta^{2}M^{2}}}\right]~,
\ee
where $\Omega_{8}$ is the volume of a unit 8-sphere
and $M=\sqrt{V''(\phi)/4 \tau_{9}}=\sqrt{|t|^{2}-1/2}$ 
is the mass of the particles in the gas
(in $\ls=1$ units).
Assume now that the temperature is small enough that 
$e^{\beta M}\gg 1$ (note that in any case 
we are interested in temperatures lower than the
Hagedorn temperature, i.e., $\beta\ge\beta_{H}=2\sqrt{2}\pi$).
Keeping only the first term in the expansion of
the logarithm and carrying out the $x$ 
integral, we are left with
\be \label{fbessel}
F(\phi,\beta)\simeq 2\tau_{9}e^{-2|t|^{2}}
-2(2\pi)^{-5}(M/\beta)^{5}K_{5}(\beta M)~.
\ee

Next, we would like to find the minimum of (\ref{fbessel}), 
i.e., the value of $\phi$ at which 
$\p_{\phi}F(\phi,\beta)=0$. It is most convenient to write this as an 
equation for the value of $M$ at the minimum,
\be \label{mineq}
\gs(2\pi)^{4}(M/\beta)^{4}K_{4}(\beta M)e^{2M^{2}+1}=4~.
\ee
To arrive at this equation we have made use of the fact that 
$\p_{x}[x^{\nu}K_{\nu}(x)]=-x^{\nu}K_{\nu-1}(x)$, and we have
introduced the string coupling constant \gs\ using 
the formula for the D9-brane tension, 
$\tau_{9}=1/\gs(2\pi)^{9}\ls^{10}$.
It can be seen from (\ref{mineq}) that in the free string limit
$\gs\to 0$ the finite-temperature effects disappear, and the minimum 
remains at $M=\infty$ ($\phi=0$). The reason for this is simply that the 
D9 tension, which sets the scale of the zero-temperature potential 
(\ref{v}), goes to infinity. We should therefore consider
\gs\ to be small but finite.

The numerical solution to (\ref{mineq}) is shown in Fig.~2. 
For $\beta,M\gg 1$ one can use the exponential approximation to the 
Bessel function to obtain the analytic expression
\be \label{minapp}
M(\beta)\simeq{\beta\over 2}+{\ln\beta\over\beta}
 -{1\over\beta}\left[\ln\gs+{9\over 2}\ln\pi-2\ln 2+1\right]~,
\ee
which gives a good approximation to the exact solution 
for all $\beta,\gs$ of interest, and makes the 
$\gs$-dependence manifest. 
As noted before, for any given $\beta$,
$M(\beta)\to\infty$ as $\gs\to 0$.  

\vspace{0.5cm}

\begin{figure}[htb]
\centerline{\epsfxsize=7cm\epsfbox{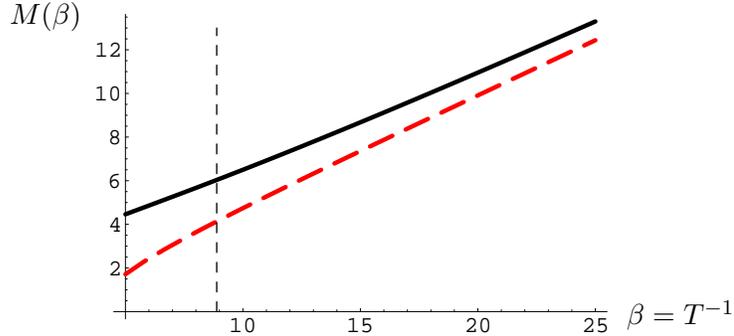}}
 \begin{picture}(0,0)
  \put(31,47){\small $M(\beta)$}
  \put(113,7){\small $\beta=T^{-1}$}
 \end{picture}
 \vspace*{-0.7cm}
\caption{\small Mass of the open string tachyon in string units 
as a function of the inverse 
temperature, for $\gs N=10^{-10}$ (solid line) and $\gs N=1$ (dashed 
line), where $N$ is the number of D9-\Dn\ pairs. The vertical line 
indicates the Hagedorn temperature, $\beta_{H}=2\sqrt{2}\pi$. See text for 
discussion.}
\end{figure}

It is seen from (\ref{minapp}) and
Fig.~2  that the mass $M$ of the 
tachyon\footnote{Strictly speaking, 
the effective mass of the tachyon is not 
really $M$ but the second derivative of (\ref{f}), and one
would ideally  carry out a self-consistent calculation using  
this mass instead of $M$ as an input to (\ref{f}). 
However, as long as $M$ is not too 
close to zero, the difference between
these two masses is small,
and so can be neglected to the accuracy of our analysis.}
is inversely proportional to the temperature, and that for
reasonably small values of $\gs$ it becomes of order one in string 
units at sub-Hagedorn temperatures 
(e.g., $M\sim 10$ at $\beta\sim 18$,
which corresponds to $T\sim T_{H}/2$). 
This is then a clear indication that the 
open string degrees of freedom 
are no longer negligible at such temperatures.\footnote{Of course, 
the fact that the theory is strongly-coupled 
near the closed string vacuum means that the free mass $M^{2}\sim V''$ 
can receive large corrections  \cite{kls}.
But even if we do not interpret $M$ as the tachyon mass, 
a low value of $M$ implies that the minimum has shifted 
significantly. Note also that, away from $|\phi|=\sqrt{\pi/2}$,
the coupling can be reduced by taking \gs\ to be small 
enough.} 
As stated before, at high enough temperatures the minimum may shift 
all the way to $\phi=0$, meaning that the open string vacuum is no 
longer tachyonic.
The relevant temperature should be comparable to 
the energy needed to completely recreate the D9-\Dn\ pair, and 
therefore (for small \gs)
larger than the Hagedorn temperature, which means it
cannot be detected directly within our present framework. We will 
return to this issue in Section \ref{stablesec},
using a slightly different approach.

 The preceding analysis has taught us that, since the tachyon field 
 has the option of uncondensing only partially, 
 one can have open strings (albeit of large mass) 
 even if the energy density is lower than what is required to 
 create a full \DD\ pair.  
 Notice that, even though closed strings are certainly 
 present as low energy 
 excitations about the $|\phi|=\sqrt{\pi/2}$ vacuum, we have been
 able to ignore them in the foregoing analysis by working
 in the canonical ensemble, where the open and closed string gases
 simply have a common temperature. 
 In the next subsection we will see that, in fact, it is more
 appropriate to use the microcanonical ensemble,
 where a given total 
 energy $U$ has to be be distributed among the components of the system in 
 such a way as to maximize the entropy. For small $U$, this means that
 the energy will go entirely to the closed string gas, the open strings 
 becoming relevant only at a rather large value of $U$. Note, however,
 that one is free to consider a state with a
 large number $N$ of brane-antibrane 
 pairs, as we will do in more detail in the next subsection. 
 In the regime $g_s \ll g_s N \ll 1$,   
 a state with \DD\ pairs and open strings
 decays only very slowly into closed strings, 
 and can be considered metastable as long as it is stable with respect to the 
 open string interactions, governed by $g_s N$. It would be interesting to investigate
 this state further using the boundary state formalism to obtain
 the supergravity fields that it generates (however on has to take into account
 the fact that the system is at finite temperature \cite{AGV}).

 Even though states of this sort have lower entropy than 
 a gas of closed strings with the same energy,  
 they could conceivably play a role in certain physical phenomena. 
 One such phenomenon 
 might be the Hagedorn transition, about which we would now like to make 
 some speculative remarks.

 Recall first that the Hagedorn temperature for closed strings is 
 `non-limiting': it can be reached by supplying a 
 finite amount of energy. Open strings, on the other hand, have a 
 limiting Hagedorn temperature (for detailed discussion on this and related 
 issues, see, e.g., \cite{hagdienes,hagbarbon}). 
 This fundamental distinction makes
 the closed string case rather mysterious,\footnote{We 
 are grateful to Bo Sundborg for emphasizing this point to us.} and 
 throughout the years several authors have speculated about the 
 possible nature of the phase transition associated
 with the non-limiting behavior (see, e.g., \cite{hag}, or the recent works 
 \cite{hagrecent} and refs. therein).
 
 If one considers string theory models of hadrons, then the Hagedorn 
 transition is naturally associated with the deconfinement transition.  
 In that context strings 
 are flux tubes of the color electric field, which disappear 
 above the critical temperature, since the flux can spread. If instead
 one considers closed strings to be fundamental objects, this picture does 
 not seem relevant. However, as discussed recently, 
 in the context of unstable brane systems
 closed strings can in fact be described as flux 
 tubes of the open string gauge field, which has been argued to be 
 confined in the closed string vacuum \cite{confinedu1,kls}.  
 If the analogy can be pushed 
 further, the Hagedorn transition could perhaps be related with 
 the spread of these flux tubes, which is only possible if D-branes 
 appear.  This would imply that for finite values of the string 
 coupling constant, the closed string spectrum changes drastically as 
 one approaches the Hagedorn temperature,
 and the closed strings ultimately 
 disintegrate into open strings, which cannot then 
 be heated above the Hagedorn temperature. 
  
 Remember also that the usual 
 analysis of string thermodynamics
 reveals that as one approaches the Hagedorn temperature the energy goes 
 into a single highly excited string, which looks like a random walk 
 that fills the entire volume available.  
 If this long string is built from electric flux, then
 it seems likely that the state can be seen as a bubble 
 where electric flux is unconfined.
 
  Unfortunately, the simple-minded 
 calculations of the present section are not enough to pursue these   
 ideas further, even if they suggest that a relation between the 
 Hagedorn transition and the emergence of open string degrees of 
 freedom is possible. Closely related remarks have been made in 
 \cite{hagbarbon}.

\subsection{Multiple brane-antibrane pairs}
\label{multiplesec}

How would the discussion in the previous subsection
change for $N>1$ D9-\Dn\ pairs? 
Strings running between the branes and the antibranes give rise to
$N^{2}$ complex tachyon fields 
which can be assembled into an  
$N\times N$ matrix $t_{a\bar{c}}$, $1\le a,c \le N$.
The potential is \cite{kmmsuper} 
\be \label{vN}
V(\phi)=2\tau_{9}\tr\exp[-2|t(\phi)|^{2}]~.
\ee
The matrices $(t\, t^{\dagger})_{ac}$ and 
$(t^{\dagger}t)_{\bar{a}\bar{c}}$ can be simultaneously
diagonalized 
by forming appropriate linear combinations of the branes and 
(independently) the antibranes, so one can restrict attention to the 
case where only the diagonal tachyons condense.  
The potential (\ref{vN})
is minimized by
\be \label{Ncond}
t_{a\bar{c}}=t\delta_{ac}, \quad t\to\infty  
   \qquad\Longrightarrow\qquad   V=0~,
\ee
which leads to a shift in the mass of all $N^{2}$ tachyons.
For our purposes, the result is to increase the
first and second terms in (\ref{fbessel})
by respective factors of $N$  and $N^{2}$, 
reflecting the energy of the branes and the
number of tachyon species:
\be \label{fbesselN}
F(\phi,\beta)=2N\tau_{9}e^{-2|t|^{2}}
-2(2\pi)^{-5}N^{2}(M/\beta)^{5}K_{5}(\beta M)~.
\ee
The net effect is simply to replace \gs\ with 
$\gs N$ in Eqs. (\ref{mineq}) and (\ref{minapp}), a fact which has
already been taken into account in the caption of Fig.~2. 

Having noted the possibility for an arbitrary number $N$ of 
brane-antibrane pairs to reemerge from the vacuum, it becomes clear that
the value of $N$ is not arbitrary and should be determined
thermodynamically. 
The $N$-dependence displayed by (\ref{fbesselN}), and 
the relative minus sign between the two terms, imply 
that a plot of the free energy as a function of $N$ has the shape of 
an inverted parabola, and is therefore minimized at $N=\infty$ for 
all $\beta<\infty$. We thus conclude that, in the canonical ensemble,
the minimum (\ref{minapp}) is unstable: the system can lower its 
free energy by increasing $N$
(with $M$ decreasing in the process), driving the 
coupling constant $\gs N$ into the non-perturbative regime.
The closed string vacuum at finite temperature
is of course already known to be unstable 
due to black hole formation, and in Section \ref{4sec} we will see 
that these two types of instabilities are in fact related (see 
\cite{hagbarbon} for related discussions).
At the same time, it is clear
that the system is not unstable in a microcanonical analysis:
given a finite amount of energy, the entropy vanishes if we
use it all up to create \DD\ pairs, or if we create no pairs at all, 
so there is an intermediate value of $N$ which maximizes the entropy. 

Our main goal thus far has been to understand the phenomenon of 
reappearance of open strings in the closed string vacuum at finite 
temperature, but it is also interesting 
to think in these terms about the reverse process. Consider $N$ 
D9-\Dn\ pairs at the open string vacuum. As we know, they are unstable and 
will decay by tachyon condensation. In order for energy to be 
conserved in the process, 
the energy gained from sliding down the tachyon 
potential must be used to produce strings,
closed or open. We can ensure that all of the energy goes into open 
strings by taking
$\gs\to 0$ while keeping $\gs N$ fixed: emission of closed strings by 
the branes is then forbidden. 
The endpoint of the condensation 
process will then consist of an open string gas on $N$ partially 
condensed branes. The final tachyon 
expectation value will lie close to $|\phi|=\sqrt{\pi/2}$ for $\gs N\ll 1$, 
and it will move towards $\phi=0$ as $\gs N$ is 
increased.
For large enough $\gs N$, then, it is conceivable 
that the tachyon will not condense at all. We will now examine this 
possibility in more detail. 

\subsection{Critical temperature} 
\label{stablesec}

For use in the next section,
we will carry out this part of the analysis for a
D3-$\overline{\mbox{D}3}$ system. We would like to 
determine the conditions under which the open string vacuum, 
$t=\phi=0$,
is a minimum of the effective tachyon potential. When this happens, 
the `tachyon' will no longer be tachyonic, so (at sub-Hagedorn 
temperatures) the massless fields 
will make the most important contribution to the free energy.
The tachyon gas in our previous calculations must consequently be 
replaced by a gas of gluons ($+$ transverse scalars $+$ superpartners).
If one starts at $\phi=0$, then
sliding down the tachyon potential (\ref{vN}) lowers the energy of 
the system, but it also gives mass to the relative $U(N)$ 
gauge fields, and so decreases the entropy of the gas. 
We are interested in establishing which of these
effects dominates. 

The free energy 
for the brane-antibrane $+$ tachyon condensate $+$ gas 
system at temperature $T=\beta^{-1}$ is given by the 
obvious modification of (\ref{f}),
\be \label{f3}
F(\phi,\beta)=2\tau_{3}\tr e^{-2|t|^{2}}
+{\Omega_{2}\over(2\pi)^{3}}c N^{2}\beta^{-4}\int_{0}^{\infty}dx\,x^{2}
\ln\left[1-e^{-\sqrt{x^{2}+\beta^{2}m^{2}}}\right]~,
\ee
where $\tau_{3}=1/\gs(2\pi)^{3}\ls^{4}$,
$m\sim|\phi|$  
is the mass given to the gluons by the Higgs effect,
and the numerical constant $c= 8+8(7/8)$ for the relevant 
8 bosonic $+$ 8 fermionic degrees of freedom (gauge field $+$ transverse 
scalars $+$ superpartners). 
Starting at $\phi=0$ and 
letting a single diagonal tachyon condense\footnote{Letting
all diagonal tachyons condense (see the discussion following 
(\ref{vN})) would yield a factor of $N$ 
in the first term of (\ref{deltaf}), but it would also give mass 
to all $N^{2}$ species of particles, 
leaving condition (\ref{Tcrit}) unchanged.}
by an amount $\delta\phi$ gives mass to $N$ out of the $N^{2}$
species of particles in 
the gas, and so changes (\ref{f3}) by
\be \label{deltaf}
\delta F= -4\tau_{3}(\delta\phi)^{2}
   +{15\over 24}N\beta^{-2}(\delta m)^{2}~,
\ee
which is positive for large enough temperature. Disregarding the
numerical constants, we thus learn that for
\be \label{Tcrit}
T\ge {T_{H}\over \sqrt{\gs N}}~.
\ee
the open string vacuum is a minimum of the free energy (equivalently, 
a maximum of the entropy for fixed total energy\footnote{As we have 
seen in the previous subsection, a microcanonical analysis is 
preferable because in the presence of a heat reservoir the system is 
unstable towards creation of an infinite number of brane-antibrane 
pairs.}),
and the brane-antibrane pairs do not annihilate. 

To arrive at  (\ref{deltaf}) we have considered the mass that the 
relative gauge fields (and scalars)
acquire due to their coupling to the tachyon, 
but we can equivalently phrase the result in the opposite direction:
the second term of (\ref{deltaf}) 
represents a mass term for $\phi$ due to a
thermal expectation value for the relative gauge fields, 
$\expec{A^{-}A^{-}}_{T}\sim N T^{2}$, corresponding to a 
mass $m_{\phi}\sim \sqrt{\gs N} T$.

It is important to note
that the regime (\ref{Tcrit}) lies in the physically accessible sub-Hagedorn
region only for non-perturbative values of the coupling, $\gs N>1$, 
where we would expect the system to have a dual
supergravity description. 
In the following sections
we will use this insight to model black branes in 
supergravity in terms of brane-antibrane systems.

\section{Entropy of Black Branes}
\label{3sec}

In the previous section we found that at a large enough
temperature there is a stable state of 
a \DD~system where a finite number of brane-antibrane pairs  
are uncondensed. In order for this temperature regime to be 
physically accessible the system must be strongly-coupled, $\gs N>1$.
It is thus natural to expect the uncondensed \DD~state to have an alternative
supergravity description, which could 
only be a black brane at the same temperature.  
In fact, it is known that in many cases \cite{hms,antientropy} 
the energy and entropy of black branes 
admit an interpretation in terms of branes and antibranes. 
In this section we explore this idea more closely for the case of 
black three-branes in type IIB string theory  
and black two- and five-branes in M-theory. 
These non-dilatonic cases have been analyzed in their near-extremal regime in 
\cite{gkp,kt}, 
and play an important role in the AdS/CFT correspondence 
\cite{malda}. The regime far from extremality 
has been considered from the point of view of Matrix theory 
\cite{bfks,matrixbh,dmrr,boostbh} 
and by means of the string/black hole correspondence
\cite{stringbh,hp} but, to the best of our knowledge, has not been discussed 
in the context of brane-antibrane models.   
We will carry out most of the discussion for
the best-understood case of D3-branes,
leaving the M2- and M5-brane cases for Section \ref{2-5sec}.

\subsection{Brane-antibrane model for black three-branes}
\label{sec:microscopic-model}

 The model that we consider is the low-energy theory on the worldvolume of a 
 system of $N$ D3-branes and \bN\ anti-D3-branes. 
 {}From the gauge theory point of view this means that 
 the original $U(\infty)\times U(\infty)$ gauge group of the full 
 theory is broken down 
 (by the expectation value of the tachyon) to $U(N)\times U(\bN)$.
 Since the numbers  $N$ and \bN\ of uncondensed branes and antibranes
 can vary, the actual values are chosen to maximize 
 the total entropy of the system for fixed charge and mass.  

 The temperature 
 is assumed (and later confirmed) to satisfy
 $\frac{1}{\sqrt{g_s N}} \ll T \ll 1$ (and similarly with $\bN$) 
 in string units. 
As seen in (\ref{Tcrit}),
the first inequality is required for stability;
the second one allows us to ignore the massive open string modes. 
For the above temperature range to exist we must have
$g_s N\gg 1$, $g_s \bN \gg 1$, so we are necessarily
in the strong-coupling regime. 
We will take $g_s\ll 1$ to suppress closed string loops. 
Under these conditions, when a brane-antibrane pair annihilates 
the energy goes to the gas of open strings 
on the remaining branes and antibranes, rather than being emitted as closed 
strings (Hawking radiation), since the latter process is disfavored for small
$g_s$.     

Since we are trying to formulate a microscopic model 
for a supergravity solution,
the restriction to strong coupling was of course expected.
Notice, however, that the situation is rather peculiar in that
there is no weakly-coupled stable vacuum:
if $g_s N < 1$, then for any physical temperature
the open string vacuum is unstable and the \DD\ pairs will condense,
driving the system towards the strongly-coupled closed string vacuum.  
This means that, unlike other cases, here we do not have the option of 
studying the system at weak-coupling and then extrapolating to 
strong-coupling.

In the absence of a weakly-coupled regime, and
knowing that the theory is not supersymmetric,
the best one can do is to use plausibility arguments to determine
the entropy. 
If the result turns out to agree with supergravity,
perhaps the only conclusion should be that supergravity predicts 
a reasonable behavior for the system at strong coupling. 
More optimistically, one could view
an eventual agreement  as evidence for the validity of the 
microscopic model. 
With these caveats in mind, we
now proceed to compute the entropy. 

 The theory on the branes is $(3+1)$-dimensional 
 $\cN=4$ super-Yang-Mills (SYM) with gauge group $U(N)$. 
 The theory on the antibranes is also $\cN=4$ SYM, with gauge group 
 $U(\bN)$, although the supersymmetry generators of the brane and 
 antibrane theories are different. 
 At weak coupling these theories would give $8N^2+8\bN^2$ massless 
bosons, $2N^2+2\bN^2$ from the gauge fields 
($A^\mu_{ac}$,$\bar{A}^\mu_{\bar{a}\bar{c}}$) 
and $6N^2+6\bN^2$ from the scalars 
($\Phi^i_{ac}$, $\bar{\Phi}^i_{\bar{a}\bar{c}}$),
plus the same number of massless fermions 
($\Lambda^\alpha_{ac}$, $\bar{\Lambda}^\alpha_{\bar{a}\bar{c}}$).
In addition, the theories are coupled together
by the fields arising from brane-antibrane strings,
which transform in the bifundamental of $U(N)\times U(\bN)$. 
These are $N\bN$ complex tachyons ($\phi_{a\bar{c}}$) 
and $16 N\bN$ massless fermions ($\Psi^\alpha_{a\bar{c}}$) coming 
from the Ramond sector.  

 At strong-coupling the counting is modified. 
 A qualitative approach (which as we will see does not give a full answer)  
is to observe that the cubic and quartic couplings give mass 
to some of the fields when others have expectation values \cite{nearext,hp}. 
If the mass is larger than the temperature, then these degrees of 
freedom are frozen and should not be counted.  
 
 Let us remind ourselves how this works 
 in the case where there are only D-branes. 
 The scalars for example get a mass from the term
\be 
 g_s \tr [\Phi^i,\Phi^j]^2 = 2 g_s \sum_{a,b,c,d=1}^N  
 \left( \Phi^i_{ab}\Phi^i_{cd} \right) \left(\Phi^j_{bc}\Phi^j_{da} - 
         \delta_{bc} \Phi^j_{de}\Phi^j_{ea}\right)~.
\ee
On dimensional grounds $\langle (\Phi^j_{ab})^2 \rangle \sim T^2$, 
so at large $N$ the last term gives a mass of order $\sqrt{g_s N} T$ 
 to all the fields. 
The same is valid for the gauge bosons and the 
fermions.\footnote{In the case of the fermions 
the relevant coupling is 
$\sqrt{g_s} \Phi^i_{ab}\bar{\Lambda}_{bc}\Gamma^i\Lambda_{ca}$. 
Since 
$\langle \tr \Phi \rangle =0 $ and $\langle \tr \Phi^2\rangle\sim 
N^2 \langle (\Phi_{ab})^{2}\rangle\sim N^2 T^2 $,
the eigenvalues $\varphi^i$ of $\Phi^i$ satisfy 
$|\varphi^i|\sim \sqrt{N} T$.}  
In the regime $g_s N \gg 1$ this mass is much larger than 
the temperature, so in principle these degrees of freedom should 
not be excited. However,
if they are not excited, then they do not get a large mass. 
For example, if only the diagonal fields (corresponding to strings going 
from one brane to itself) acquire expectation values,
then the off-diagonal fields are given a mass of order 
$m\sim T$, which is not large enough for them to be suppressed. 
This contradiction means that the naive
argument does not provide a definite answer, 
but it does suggest that the situation is intermediate 
between these two extremes. 
At present, our limited understanding of this case 
comes from the AdS/CFT correspondence \cite{malda},
which predicts that the entropy at $g_s\ll 1 \ll g_s N$ 
is the same as that of a 
free field theory with only $6N^2$ bosons and the same number of
fermions, instead of the naive $8N^2$ \cite{gkp,threequarters}.  

 Going back to the case of branes and antibranes, 
 the situation is similar. We can apply the above argument to show that, 
 if $\cO(N^2)$ fields 
are excited on the branes and $\cO(\bN^2)$ on the antibranes, 
then the tachyons $\phi_{a\bar{c}}$ 
and the fermions $\Psi_{a\bar{c}}$ acquire a mass of order 
$\sqrt{g_s N} T$,
which is in fact what we obtained (for the tachyon) in  Section 
\ref{stablesec}. If such is the case, 
the fact that (for $g_s N\gg 1$) the temperature is much lower than 
their mass means that these fields are not excited,
and the theory on the branes decouples
from the theory on the antibranes. 
For these two decoupled theories, 
the AdS/CFT correspondence predicts that indeed $\cO(N^2)$ and
$\cO(\bN^2)$ fields are excited, respectively, 
making the result self-consistent. 

 The upshot of this long discussion is
 that the energy, entropy and charge of our microscopic system
 should be given by
\bea
 \Et &=&  (N+\bN)\tau_{3} V 
   + a {\pi^2\over 16}N^2 V T^4 
   + a {\pi^2\over 16} \bN^2 V \bT^4~, \label{eq:Eft}\\    
 \Sm &=& a{\pi^2\over 12}N^2 V T^3 + a{\pi^2\over 12}\bN^2 \bT^3~, 
      \label{eq:Sft}\\
 Q &=& N - \bN~,  \label{eq:Qft}
\eea
 where $\tau_{3}=1/(2\pi)^{3}\gs\ls^{4}$ denotes the tension of the D3-branes.
 The numerical constant $a=8$ if we consider $8$ free bosons and $8$ 
 free fermions, but $a=6$ if we use the strong-coupling
 input from the AdS/CFT correspondence \cite{gkp,threequarters}, 
 as we will do from now on. 
  We have taken the theory to be compactified on a torus of volume $V$. 
 The above formulas would be expected to hold only for $T>V^{-1/3}$, but
 in fact they are valid even for much smaller temperatures because 
the branes can be multiply-wrapped, which effectively means the gases 
 live in a much larger volume \cite{ms,hp}. 
 Finally, notice that since the theories are decoupled, 
 the gas on the branes and the gas on the antibranes 
 could a priori have different temperatures, $T\neq\bT$. 

 The energy (\ref{eq:Eft})
 includes a contribution from the brane/antibrane tension 
 (or equivalently, from the tachyon potential) and another from the 
 gases.
 These two contributions
 can be distinguished by considering the full energy-momentum tensor,
 because the tachyon potential contributes with a minus sign 
 to the space-space components of $T_{\mu\nu}$:
\bea
 T_{00} &=& (N+\bN)\tau_{3} 
    + a {\pi^2\over 16}N^2  T^4 + a {\pi^2\over 16}\bN^2  \bT^4~, 
                   \label{eq:T00ft}\\
 T_{ij} &=& \left[-(N+\bN)\tau_{3} 
    + a {\pi^2\over 48} N^2  T^4 
    + a {\pi^2\over 48} \bN^2  \bT^4 \right] \delta_{ij}~, 
\label{eq:Tijft}
\eea 
where $i,j=1,2,3$ and we have used the fact that the 
trace of $T_{\mu\nu}$ is zero for a gas of massless particles. 
So, for example, 
one can separate the energy due to the brane tension by computing
 $T=\eta^{\mu\nu} T_{\mu\nu}$, which receives
 no contribution from the gases. 
Expressions (\ref{eq:T00ft}) and (\ref{eq:Tijft})
are useful to us because the same logic can be applied in the 
supergravity side: $T_{\mu\nu}$ can be computed 
using the asymptotic value of the black brane metric, 
and then both contributions can be distinguished also in that case.

 The last step in the formulation of the microscopic model
 is to maximize $\Sm$ with respect to $N$ and $\bN$,
 keeping $\Et$ and $Q$ fixed. In the next two subsections
 we will do this and then compare the result with supergravity.

\subsection{Comparison with supergravity: neutral case}
\label{sec:comp-with-supergr}

We consider here the neutral case, $Q=N-\bN =0$, leaving
the more general case to the next subsection. 
Since $N=\bN$, by symmetry we expect the temperatures
to be equal, $T=\bT$. 
  Thus, the energy and entropy are given by
\bea
 \Et &=& 2N\tau_{3}V + a{\pi^2\over 8}N^2 V T^4~,  
       \label{eq:EQ=0}\\
 \Sm &=&a{\pi^2\over 6} N^2 V T^3~.  \label{eq:SQ=0}
\eea
As discussed in the previous subsection, including strong-coupling effects
we expect $a=6$. 
If one is not interested in the 
numerical coefficient then $a$ can be taken as an unknown constant,
and the correct functional form of the entropy should follow just 
from the fact there are $\cO(N^2)$ massless degrees of freedom. 

 The value of $N$ is determined by maximizing the entropy at fixed 
 $\Et$.  {}From the above relations we obtain
\be
\Sm = a^{\frac{1}{4}}\frac{2^{\frac{5}{4}}}{3} \sqrt{\pi} \sqrt{N} 
       V^{\frac{1}{4}} \left(\Et - 2NV\tau_{3} \right)^{\frac{3}{4}}~,
\label{eq:S2Q=0}
\ee
which is maximized by
\be
N = \frac{1}{5} \frac{\Et}{\tau_{3}V}~.
\label{eq:NQ=0}
\ee
 This is easily seen to imply that the energy contained in 
 the gases 
 is $3/2$ of the total tension of the branes and antibranes,
 a prediction which as is shown below agrees with supergravity. 
 Plugging the equilibrium value (\ref{eq:NQ=0}) back
 into (\ref{eq:S2Q=0}) we obtain the entropy-energy relation
\be
\Sm = a^{\frac{1}{4}}2^{\frac{5}{4}} 3^{-\frac{1}{4}} 5^{-\frac{5}{4}}   
      \pi^{\frac{1}{4}} \sqrt{\kappa}  
      V^{-\frac{1}{4}} \Et^{\frac{5}{4}}~,
\label{eq:SEftQ=0}
\ee
where we have expressed the D3-brane tension in terms
of the gravitational coupling constant, $\tau_{3}=\sqrt{\pi}/\kappa$.
In addition, we find that 
the temperature is $T\sim (g_s N)^{-\frac{1}{4}}$,
which as required satisfies $\frac{1}{\sqrt{g_s N}} \ll T \ll 1$.  

 To compare with supergravity, we recall that a neutral black three-brane 
 with Schwarzchild radius $r_0$ has mass and entropy \cite{entropy} 
\bea
\Ms &=& \frac{5}{2}\frac{\pi^3}{\kappa^2}r_0^4 V~,\\
\Ss &=& \frac{2\pi^4}{\kappa^2}r_0^5 V~,
\eea
which implies that
\be
\Ss = 2^{\frac{9}{4}} 5^{-\frac{5}{4}} \pi^{\frac{1}{4}}  
    \sqrt{\kappa} V^{-\frac{1}{4}} \Ms^{\frac{5}{4}}~.
\ee
 Identifying $\Ms=\Et$, we see that the functional form of the 
 supergravity and field theory entropies agree.  
 The numerical coefficient would agree for $a=48$, but 
 as we have seen the AdS/CFT correspondence implies that $a=6$,
 meaning that the field theory entropy is a factor of $2^{3/4}$ too small, 
 $\Ss=2^{3/4}\Sm$. Equivalently, one can say that the supergravity 
 entropy behaves as if the gases 
 carried twice the available energy.

 A similar factor was found already 
 in \cite{hms} for the D1/D5 system far away from extremality. 
 The authors observed that 
 considering separate gases to account for the entropy results 
in an energy $4$ times larger than the correct value. 
We will see in Section \ref{2-5sec}
that the same factor appears in the M2- and M5-brane cases
after one takes into account the AdS/CFT prediction for the 
strong-coupling behavior of the worldvolume theory. 
 
 Another interesting check is to consider the energy-momentum tensor.  
 {}From the asymptotic value of the 
gravitational field one finds that \cite{dmrr} (see also the Appendix): 
$T_{ij} = -\frac{1}{5}T_{00} \delta_{ij}$. 
Putting $T_{00}=\Tau + \Eg$ and 
$T_{ij} = (-\Tau + \frac{1}{3} \Eg)\delta_{ij}$, where 
$\Tau$ is the contribution from the brane/antibrane tension 
and $\Eg$ from the gases 
(see the previous subsection), 
one obtains $\Eg = \frac{3}{2} \Tau$, in 
agreement with the field theory prediction. 

 Before moving on to the charged case, let us discuss an 
 interesting issue that appears already here. 
 Since we reproduce the black brane entropy, 
it is clear that the specific heat of our system is negative. 
To understand why, let us consider how Hawking evaporation proceeds in this 
model, and check that the temperature increases when energy is radiated. 
When a closed string is emitted,
energy is taken from the open string gas, 
so a priori the temperature 
should decrease. However, we have found that, in equilibrium, 
the energy in the gas is $3/2$ the tension of the \DD\ pairs. This 
means that, when the gas 
has lost enough energy so as to match $3/2$ the tension 
of $N-1$ pairs, it will be entropically favorable
for one pair to annihilate, giving energy to the gas 
and increasing its temperature. 
This is repeated again and again,
effectively increasing the temperature on average as the mass of the 
system decreases. 
The process will continue until $g_s N\sim 1$, 
where the gas reaches the Hagedorn temperature ($T\sim 1$ in string 
units) and the model (\ref{eq:Eft})-(\ref{eq:Qft}), 
based only on the massless open string modes,
ceases to be valid.  At this point we would expect all of the 
available energy to go into a highly-excited (open or closed)
long string, so the brane-antibrane model makes
contact with the string/black hole correspondence
\cite{stringbh,hp}. Indeed, $g_s N\sim 1$ is precisely the 
correspondence point, where the curvature of the black brane at the 
horizon is of order the string scale, and the Bekenstein-Hawking 
entropy is known to match the entropy of a long string at the 
Hagedorn temperature \cite{hp}.

\subsection{Comparison with supergravity: general case}

We start by considering the supergravity
expressions, following a procedure similar to \cite{hms}. 
 The energy, entropy and charge of a non-extremal three-brane are given by 
 \cite{entropy} 
\bea
 M &=& \frac{\pi^3}{\kappa^2} r_0^4 V \left(\cosh2\alpha 
        + \frac{3}{2}\right)~,\label{eq:Msgr} \\
 S &=& \frac{2\pi^4}{\kappa^2} r_0^5 V\cosh\alpha~, \label{eq:Ssgr}\\ 
 Q &=& \frac{\pi^{\frac{5}{2}}}{\kappa} r_0^4 \sinh2\alpha~. \label{eq:Qsgr} 
\eea
 As discussed before, it is interesting to consider not only the mass 
 but also the other components of the energy-momentum 
 tensor. As shown in the Appendix, it turns out to be
\bea
T_{00} &=& \frac{\pi^3}{\kappa^2} r_0^4 \left(\cosh2\alpha 
        + \frac{3}{2}\right)~, \label{eq:T00sgr} \\
T_{ij} &=& \left[\frac{\pi^3}{\kappa^2} r_0^4 \left(-\cosh2\alpha 
        + \frac{1}{2}\right)\right] \delta_{ij}~.\label{eq:Tijsgr}
\eea
  Comparing (\ref{eq:Msgr}), (\ref{eq:Qsgr}) and (\ref{eq:Tijsgr}) with 
  (\ref{eq:Eft}), (\ref{eq:Qft}) and (\ref{eq:Tijft})  uniquely 
determines
\be
 N  = \frac{\pi^{\frac{5}{2}}}{2\kappa} r_0^4 e^{2\alpha},\ \ \ 
 \bar{N} = \frac{\pi^{\frac{5}{2}}}{2\kappa} r_0^4 e^{-2\alpha}~. 
\label{eq:NbN}
\ee
 The energy of the gases 
 is then identified with
\be
\Eg = \frac{3}{2} \frac{\pi^3}{\kappa^2} V r_0^4~,
\ee
 in terms of which the entropy can be written as
\be
 \Ss=2^{\frac{5}{4}} 3^{-\frac{3}{4}} \pi^{\frac{1}{2}} V^{\frac{1}{4}} 
     \Eg^{\frac{3}{4}} \left(\sqrt{N}+\sqrt{\bN}\right)
\ee
 {}From (\ref{eq:Eft}) and (\ref{eq:Sft}) with $a=6$, 
 we see that this is the correct entropy for a gas of particles 
 on the $N$ branes
plus another gas on the $\bN$
antibranes, both with the same energy $\Eg$. 
However, since the total energy available for the gases is $\Eg$, in the 
field theory model we have to assign an energy $\Eg/2$ to each of 
them, resulting in a mismatch in the entropy 
which is exactly the same as found for the neutral case in the
previous subsection: $\Ss=2^{3/4}\Sm$. 
Under the condition
that the energies in both gases are the same,
one can 
check that the expressions (\ref{eq:NbN}) 
are the ones that maximize the 
entropy for fixed charge and mass. 
 
 The fact that the energy densities (or equivalently the pressures) of 
 the two gases are the same implies that their temperatures $T$,$\bT$ 
 are different. They are related through
\be \label{2temps}
\frac{2}{\Th} = \frac{1}{T} + \frac{1}{\bT}~,
\ee
where $\Th$ is the temperature defined as 
$\Th^{-1}=(\partial S/\partial M)_{Q}$. Due to the discrepancy
in the numerical coefficient of the entropy, \Th\ is a factor of
 $2^{3/4}$ smaller than the
Hawking temperature of the black brane, $T_{h}=1/\pi r_o\cosh\alpha$. 
An expression of the same form as (\ref{2temps}) 
appeared in the analysis of the D1/D5 system, 
relating the Hawking temperature to the
temperatures of the left- and right-moving modes 
in the microscopic description \cite{nearext}. 
Remarkably, in that case the two distinct temperatures 
could be recognized also in the supergravity side, by computing the 
greybody factor \cite{msgrey}. It would be interesting to see whether 
the same can be done here, examining the absorption probabilities 
for the non-extremal three-brane. For the near extremal case
the absorption probability was computed in \cite{d3grey} and also in 
\cite{PSabs} where a much more precise analysis was done. It would be 
interesting to extend the results to the region far from extremality. 

 As we emphasized before, since the theories on the branes and 
 antibranes are decoupled, there is nothing to prevent us from
 postulating that the temperatures of the 
 corresponding gases are different.
 However, it is not clear to us why supergravity seems to require that
 their energies be the same.\footnote{A similar correlation between 
 the various components of the system was observed in \cite{hms}. 
 It is intriguing to note the similarity between these correlations 
 (and their accompanying numerical discrepancies)  
 and certain 
 aspects of the field theory model for eternal AdS black holes that was 
 very recently formulated in \cite{juan}.}
 It is conceivable that states in which the gases 
 have different energies (or equivalently, different pressures)
 are allowed microscopically. If so, they could correspond to
 other supergravity backgrounds 
 (perhaps like the ones found in \cite{bmo}),
 or they might not even have a supergravity description.
 A clue in this respect might be the observation that, under this condition, 
 the fluctuations in the transverse positions of the branes and the 
 antibranes turn out to be equal, and of the order of the 
 Schwarzschild radius. 
 This follows from the fact that such fluctuations are measured 
 by the eigenvalues of the $N\times N$ scalar field matrix $\Phi^i_{ac}$, 
($i=1,\ldots,6$). As noted in subsection ~\ref{sec:microscopic-model}, 
the eigenvalues are of order $N T^2$ for the branes and $\bN \bT^2$ 
for the antibranes, with agreement if
both gases have the same energy density ($N^2 T^4 = \bN^2 \bT^4$). 

The equality of the two energies implies that,
as one approaches the extremal limit $M= Q\tau_{3}V$ 
(i.e., $\alpha\to\infty$ with $r_{0}\propto e^{-\alpha/2}$), 
the temperature of the gas on the antibranes 
grows without bounds. On the other hand,
since the model is based on massless open string modes, it
is expected to be valid only if both $T$ and $\bT$ are
substantially lower than the Hagedorn temperature $T_{H}\sim 1/\ls$. 
{}From the microscopic 
perspective, we know that the 
Hagedorn temperature is in this case limiting
(see, e.g., \cite{hagdienes,hagbarbon}), and we expect
that as $\bT\to T_{H}$, the 
energy available to the  gas on the antibranes
goes to a highly excited long string, whose contribution to the 
entropy is however negligible compared to the gas on the branes. 

In the near-extremal region, it is known that the black brane 
entropy can be precisely reproduced using a 
system without antibranes \cite{gkp}, so as one approaches extremality
one would intuitively expect a transition to this class of states. 
It is easy to see that the entropies of these two 
microscopic descriptions cross when $M\simeq 6 Q\tau_{3}V$, which 
indeed suggests a transition, with the 
brane-antibrane system being the preferred one further away from 
extremality.
However, it is hard to see how one could retain the agreement with 
supergravity in the entire parameter space: the brane-antibrane model 
gives the exact dependence of the black brane entropy on $M$ and $Q$, 
but gives a numerical value which is
always a factor of $2^{3/4}$ too small, whereas the model of 
\cite{gkp} is in accord with supergravity in the 
near-extremal region, but deviates significantly from it
already at the presumed transition point.
To summarize, at least from the microscopic point of view
there appears to be a discontinuity in taking the near-extremal limit, 
but the issue clearly deserves futher study.

\subsection{Two- and five-branes in eleven dimensions}
\label{2-5sec}

The calculation of the entropy can easily be generalized to the case of M2- 
and M5-branes. 
For the neutral case, 
our microscopic model is again of the form 
\be
\Et=2N\tau _{p}V+aN^{c}VT^{p+1}~,
\ee
with the entropy given by 
\be
\Sm=\frac{p+1}{p}aN^{c}VT^{p}~,
\ee
where $c$ and $a$ are numerical constants 
which depend on the type of brane. They 
follow from the near-extremal case \cite{kt} or equivalently,
from the AdS/CFT correspondence.
Proceeding in the same way as in the case of the D3-brane one finds 
\be
\Sm=2^{-\frac{c}{p+1}}a^{\frac{1}{p+1}}\tau _{p}^{-\frac{c}{p+1}}
p^{-\frac{1}{p+1}}\left( c+p\right)^{\frac{c-1}{p+1}}
c^{-\frac{c}{p+1}}V^{\frac{1-c}{p+1}}M^{\frac{c+p}{p+1}}~.
\ee

In the case of the M5-brane the tension is given by 
\be
\tau_{5}=2^{-1/3}\pi ^{1/3}\kappa ^{-4/3}~,
\ee
and the calculations for the near-extremal case suggest that 
\be
a=2^{7} 3^{-7}5\pi^{3},\qquad c=3~.
\ee
This gives an entropy
\be
\Sm=2^{-13/6}3^{1/3}\pi ^{1/3}\kappa ^{2/3}V^{-1/3}M^{4/3}~,
\ee
which has the right dependence on the mass and volume as compared with the 
supergravity result \cite{entropy}, 
but is a factor $2^{5/6}$ too small.

For the M2-brane the tension is given by 
\be
\tau_{2}=2^{4/3}\pi^{2/3}\kappa^{-2/3}~,
\ee
and the near-extremal case suggests that 
\be
a=2^{11/2}3^{-4}\pi ^{2},\qquad c=\frac{3}{2}~,
\ee
so one finds 
\be
\Sm=2^{3/2} 3^{1/6}7^{-7/6}\pi ^{1/3}\kappa ^{1/3}V^{-1/6}M^{7/6}~,
\ee
which again has the right behavior but this time is a factor
$2^{2/3}$ too small.
It is intriguing that the numerical
factor can be corrected in the same way for all
three models: the  $2^{p/(p+1)}$ discrepancy means that 
in the three cases the supergravity entropy behaves as if the gas 
carried twice the available energy.

The charged case is again similar to the D3-brane described in the 
previous subsection. The supergravity formulas can be found in 
\cite{entropy}.
The black five-brane solution in eleven dimensions has
\bea
\Ms &=& \frac{2\pi^2}{\kappa^2} r_0^3 V 
    \left(\cosh2\alpha + \frac{5}{3}\right)~, \\
T_{ij} &=& \frac{2\pi^2}{\kappa^2} r_0^3  
    \left(-\cosh2\alpha + \frac{1}{3}\right) \delta_{ij}~,\\
\Ss &=& \frac{16\pi^3}{3\kappa^2} r_0^4 V\cosh\alpha~, \\
Q &=& \frac{2^{\frac{3}{2}}\pi^2}{\kappa^{2/3}} r_0^3 \sinh2\alpha~,
\eea
while for the black two-brane one has 
\bea
\Ms &=& \frac{\pi^4}{2\kappa^2} r_0^6 V 
    \left(\cosh2\alpha + \frac{4}{3}\right)~, \\
T_{ij} &=&  \frac{\pi^4}{2\kappa^2} r_0^6  
   \left(-\cosh2\alpha + \frac{2}{3}\right) \delta_{ij}~,\\
\Ss &=& \frac{2\pi^5}{3\kappa^2} r_0^7 V\cosh\alpha~, \\
Q &=& \frac{\pi^4}{\sqrt{2}\kappa^{4/3}} r_0^6 \sinh2\alpha~.
\eea
It is easy to see that these formulas can again
be interpreted in  
terms of a brane-antibrane model with
\bea
\Et &=& (N + \bN) \tau_p V + a N^c V T^{p+1} + a N^c V \bT^{p+1}~, \\
T_{ij} &=& \left[-(N + \bN) \tau_p V 
    + \frac{1}{p} \left(a N^c V T^{p+1} 
    + a N^c V \bT^{p+1}\right)\right]\delta_{ij}~, \\
\Sm &=& \frac{p+1}{p} \left(a N^c V T^{p} + a N^c V \bT^{p}\right)~, \\
Q &=& N - \bN~,
\eea
and that the inferred values of $N$ and $\bN$
correctly maximize the entropy for a given mass and charge. 
Just like in the D3-brane case, one finds here that, to match 
the supergravity expressions,
the energy densities of both gases have to be the same. 
The constant $c$ turns out to be the same as 
in the near-extremal case \cite{kt}, 
but $a$ differs by a factor of $2^p$ as before.

\section{Instabilities of Black Branes}
\label{4sec}

 As shown by Gregory and Laflamme \cite{GL}, black $p$-branes are generally 
 unstable, and tend to split into lower-dimensional branes. 
 This can be seen thermodynamically, 
by showing that lower-dimensional branes of the same mass 
have larger entropy, or
dynamically,  by studying metric perturbations. 
The best studied example is the black string,
which is believed to collapse into a series of black holes (see 
however, the recent paper \cite{hm}). 
If the black string is charged, 
one expects the black holes to be threaded by an
extremal (or near-extremal) black string carrying the charge. For 
higher-dimensional branes one has similar expectations. 
That the thermodynamical and the dynamical instabilities 
are related is not obvious. Recently 
it has been conjectured and checked in several examples that the 
classically unstable mode appears when the entropy is not a local minimum
\cite{GLthermo}. 
This is another remarkable example of the deep connection between 
gravity and thermodynamics.

 An important property of the Gregory-Laflamme transition is that it does not
occur if the black brane is wrapped on a torus of sufficiently small 
size,
as can be seen by examining the expressions for the entropy.
For simplicity, we will restrict attention to the case of a brane 
with no charge. 
The entropy of a neutral black $p$-brane in $D$ dimensions is given by 
\begin{equation}
S_{p}=2\pi \Omega _{n+1}^{-\frac{1}{n}}
    \left( \frac{2}{d+1}\right)^{\frac{n+1}{n}}
    \kappa^{\frac{2}{n}}V_{p}^{-\frac{1}{n}}M^{\frac{n+1}{n}}~,
\label{sugrent}
\end{equation}
where $n=D-p-3$, $\Omega _{n+1}$ is the volume of a unit 
$(n+1)$-sphere, and $V_{p}$ is the volume of the $p$-dimensional torus 
on which the brane is wrapped. If
the radii of the torus are small enough the black $p$-brane is, according to
the above reasoning, stable, while if one of the radii, $R$, is too large
the black $p$-brane will tend to collapse in the corresponding direction
forming a black $(p-1)$-brane. For this to occur the lower-dimensional brane
should be the configuration with the highest entropy. Ignoring numerical
factors, we find from setting $S_{p}\sim S_{p-1}$ that the critical value of 
$R$ is given by 
\[
R\sim \left( \frac{\kappa^{2}M}{V_{p-1}}\right)^{\frac{1}{n+1}}, 
\]
which coincides with the Schwarzschild radius 
(or equivalently, the inverse Hawking temperature) of the $(p-1)$-brane. 
If the conjectured
equivalence between the dynamical and thermodynamical 
instabilities is correct, we conclude that the $p$-brane is
stable only if \emph{all} radii are smaller than this critical value, 
i.e.,
\be \label{rcrit}
R_{i}<  l_{P}(M l_{P})^{1\over D-3} \qquad\forall\quad i=1,\ldots,p~,
\ee
where $\lP\sim \kappa^{2/(D-2)}$ is the  $D$-dimensional
Planck length.

Can the properties of the Gregory-Laflamme transition be understood in 
terms of our brane-antibrane construction? It is clear that, since we have 
the correct formulas for the entropy, the thermodynamical instability is 
exactly the same. {}From this point of view, the instability
means that if we enclose the 
system in a finite volume and hold its total energy fixed, then
its entropy will increase when  the volume decreases. 
In other words, the system prefers not to occupy the entire volume 
available, and it can accomplish this
by annihilating the \DD\ pairs in one region of 
space to create additional pairs in another. 
This is not in contradiction with the fact that, as we discovered in 
Section \ref{stablesec}, at the  temperatures of interest
the `tachyon' has a positive mass-squared,
and the open string vacuum is stable. 
The analysis there was restricted to the case of a constant 
tachyon, whereas here we are considering
a space-dependent perturbation which 
can be described 
as a local variation in  the number of branes and antibranes. 
Such perturbation can be unstable even if the 
constant one is not. 
It would be interesting to pursue this issue further and find
the dynamical equations of the perturbation to obtain the unstable mode 
analogous to the one found in supergravity by Gregory and Laflamme 
\cite{GL}. 
Here, however, we limit ourselves 
to a simpler but nevertheless interesting question. 
We have noted that the entropy 
increases as we reduce the volume, but if we reduce the volume too 
much, the temperature might not be high
enough to excite the gas, and the entropy would 
in fact decrease. It is then conceivable that the microscopic model 
stabilizes itself at some finite volume. 
Since we keep the total energy fixed, and not the temperature, 
it is not clear a priori whether this happens or not.
 We will analyze this issue in the next subsections,
 and in the process obtain a simple model for a
ten-dimensional 
black hole in terms of branes and antibranes.
 The fact that certain 
aspects of the Gregory-Laflamme instability 
could be explained in terms of a D-brane model
was noted already in \cite{hp}. 

\subsection{Collapse into a black hole}
\label{bhsec}   

We  consider now the case where all of the cycles of the torus violate 
the bound (\ref{rcrit}), and so, for the purposes
of this section, can be considered to be infinite. 
The brane is then unstable, and is expected to reduce
to a black hole.
Our microscopic model tells us that the mass of the neutral black 
brane can be written as 
\be \label{micrmod}
M=2\tau_{p}NL^{p}+\Eg, 
\ee
where $\Eg$ is the energy of the gas and $L$ is the size of the brane system,
which can vary due to tachyon condensation. 
Both $\Eg$ and the entropy $S$ 
are proportional to $N^{c}$, where 
$c=2,3/2,3$ for the case of D3-, M2-, and M5-branes, respectively.

If we regard $M=M(N,L,T) $ and $S=S(N,L,T)$, then
maximizing the entropy while holding the mass fixed
means enforcing the two conditions 
\bea \label{maxn}
\left(\frac{\p S}{\p N}\right)_{M,L} &=&\frac{\p S}{\p N}
            -\frac{1}{T}\frac{\p M}{\p N}=0~,  \\
\left(\frac{\p S}{\p L}\right)_{M,N} &=&\frac{\p S}{\p L}
            -\frac{1}{T}\frac{\p M}{\partial L}=0~, \label{maxl}  
\eea
where the second one is the condition of vanishing pressure.
The first equality in (\ref{maxn}) and (\ref{maxl}) follows from the 
chain rule and the fact that (\ref{micrmod}) implies
\be \label{dsdm}
\left({\p S \over \p M}\right)_{N,L}=\left({\p S \over \p E}\right)_{N,L}
      ={1\over T}~.
\ee
With the help of (\ref{micrmod}) and $S,E\propto N^{c}$, 
we can rewrite (\ref{maxn}) as 
\be  \label{dsdn}
cS-\frac{M}{T}+\frac{1-c}{T}E=0~.
\ee
By dimensional analysis, $S\sim S\left( LT\right)$, and
$E\sim f\left( LT\right)/L$ for some function $f$. Using this and          
$M-E\propto L^{p}$, together with (\ref{dsdm}) and (\ref{dsdn}), 
we can rewrite the first equality in (\ref{maxl}) as 
\begin{equation} \label{dsdl}
\left(\frac{\p S}{\p L}\right)_{M,N}=
    \frac{1}{TL}\left(\left( p+1\right) E-pM\right) .
\end{equation}

Now, for a large $p$-brane the equation of state for the gas is 
$S=\frac{p+1}{p}\frac{E}{T}$. 
With the help of (\ref{dsdn}) and (\ref{dsdl}), this implies that the 
system has negative pressure, 
\be
\left(\frac{\p S}{\p L}\right)_{M,N}=
     -\frac{\left( c-1\right) p}{p+c}\frac{M}{TL}<0~, 
\ee
with the consequence that the branes will tend to contract. 
That is, Eq.~(\ref{maxl})
can never be satisfied: as we knew already,
there is no local minimum. 
We would now like to consider how finite-size effects
change the form of the entropy. The
partition function for the gas (including the bosonic and fermionic degrees 
of freedom) is given by 
\be
\ln Z\sim N^{c}\sum_{\vec{n}\in\mathbf{Z}{}^{p}}
   \ln\left(\frac{e^{|\vec{n}|/LT}+1}{e^{|\vec{n}|/LT}-1}\right)~, 
\ee
When the size becomes smaller than $1/T$, the partition function
is dominated by the $|\vec{n}|=1$ term, 
\be
\ln Z\sim N^{c}e^{-1/TL}~, 
\ee
{}from which one easily derives 
\be
E\sim \frac{N^{c}}{L}e^{-1/TL} 
\ee
and 
\begin{equation}
S=LE+\frac{1}{T}E~.  \label{bhent}
\end{equation}
The crucial question is now whether this change will be sufficient to stop
the contraction. For this to occur the pressure (\ref{dsdl}) must  
vanish, i.e.,
\be \label{bhen}
E=\frac{p}{p+1}M~. 
\ee
Using (\ref{dsdn}) and (\ref{bhent}) one finds that this will indeed 
happen at a critical size 
\begin{equation}
L=\frac{1}{pc}\frac{1}{T}~.  \label{bhsize}
\end{equation}

To summarize, we have shown that a black brane wrapped on a large torus is
unstable, and will tend to break and contract until it reaches the size
quoted above. What are the characteristics of the final object? 
Using (\ref{bhen}) and (\ref{bhsize}) in (\ref{bhent}) we find that 
$c(p+1)S=(pc+1)M/T$, and with the help of (\ref{dsdm}) 
we can infer from this that 
\be
S\sim M^{\frac{c\left( p+1\right) }{pc+1}}. 
\ee
More precisely, if we use the
values appropriate for the D3-brane case, 
$p=3$ and $c=2$, we find  
\be \label{bh10}
S\sim (M l_{P,10})^{8/7}~, 
\ee
which is the expected entropy for a ten-dimensional black hole! 
The M2-brane, with $p=2$ and $c=3/2$, and the M5-brane, 
with $p=5$ and $c=3$, 
both give the appropriate entropy for an eleven dimensional black hole: 
\be \label{bh11}
S\sim (M l_{P,11})^{9/8}~. 
\ee

Notice also that the size of the 
final object is commensurate with the Schwarzschild radius $r_{0}$ of the 
corresponding black hole: one finds (in Planck units)
$L\sim M^{1/7}$ for the D3-brane case and 
$L\sim M^{1/8}$ for the M2- and M5-brane 
cases. By (\ref{bhsize}), this is equivalent to saying that the 
temperature of the gas agrees with the Hawking temperature of the 
black hole.
It is intriguing 
that in the microscopic model $r=r_{0}$ has a 
definite meaning as the place where the system ends; 
it is difficult to imagine how an infalling observer could fail to notice 
this type of `horizon', as must be the case for a large macroscopic black hole. 
 
All in all we have obtained a simple model for the ten-dimensional 
 black hole.  
It is perhaps worth noticing that the expression (\ref{bh10}) 
 can be writen as $S\sim N^2$,
that is, the entropy is proportional to the square of the number of 
branes-antibrane pairs, and depends on the volume
and temperature only through $N$. 
This comes from the fact that there are $N^2$ modes with fixed 
spatial dependence ($|\vec{n}|=1$).  
This is similar to what happens with a `small' black hole in
global AdS space. 
{}From there we know that a ten-dimensional black hole 
can be described by constant matrices
of size $N\times N$ (see \cite{flatads}), 
which is quite similar to what we have here. 
 It would be interesting to further
explore the properties of the present model,
to see if they agree with the expectations 
from supergravity.

 We have so far shown that a $p$-brane wrapped on a large torus is unstable, 
 and that the end product will be a stable black hole, 
 in complete agreement with
the supergravity expectations. 
What if the size $R$ of the torus is small? We then
need to show two things:
that the standard formula for the entropy of the brane 
continues to hold for the small tori, and that
there is no instability towards a
collapse into something else. 
The crucial observation in this context is
that the branes can connect with one another to form a single
multiply-wrapped brane \cite{ms,hp,bfks}. 
Effectively this means fewer degrees of freedom living in a
larger volume. In the D3-brane case, for instance,
we find $N$ degrees of
freedom, rather than $N^{2}$, living in the volume $NV$, rather than $V$. 
For large $V$ this implies that the formula for the entropy remains exactly the
same whether the branes are multiply-wrapped or not. For small $V$, however,
there is an important difference:  in the multiply-wrapped
case the entropy formula continues to
be of the standard form down to much smaller size. 
The reason is of course that the effective volume is really $NV$, and
as a consequence finite-size effects are delayed. 
This is crucial since otherwise
we would not have been able to explain the entropy of the black brane for
small tori, which is precisely where the system should be stable. 
Furthermore,
ignoring the Gregory-Laflamme instability, 
the supergravity expression for the entropy
of a $p$-brane does not show any discontinuity at the the critical size.

Still, we must make sure that the instability that we observed for large tori
is no longer present when $R<1/T$. 
The key is again the multiple wrapping. 
If the tachyon
condenses leaving a bubble of uncondensed tachyons of size $L<R$, 
it is clear that within this bubble we cannot have any multiply-wrapped branes. 
Since $L<1/T$, finite-size effects are important, 
and as we have seen above these would
yield the smaller entropy (\ref{bhent}). 
{}From this we conclude that, on a small torus, the (multiply-wrapped) 
brane-antibrane system is stable, 
in agreement with the expectations for its supergravity counterpart. 
A subtlety is that it might be possible for the tachyon
to condense in such a way as to create a hole on the branes and 
antibranes, without disrupting the fact that they are multiply-wrapped.
To probe the stability of the system against such local perturbations
would require a more detailed analysis of the 
dynamics than what we consider in this paper.

\subsection{Collapse into lower-dimensional branes}

What if we let only some of the cycles of the $p$-dimensional torus
on which the system lives be larger than the bound 
(\ref{rcrit})?  Can our model reproduce the entropies of the resulting 
lower-dimensional black branes?  To be more precise, let us say that 
$p-\tilde{p}$ of the radii of the $\mathbf{T}^{p}$
exceed the stability bound. 
The \DD\ pairs will then 
partially annihilate, and the size $L$
of the system along these directions will decrease.  
The entropy increases if we decrease $L$, but as we show below 
this ceases to be true when $L\sim 1/T$.  
Since our model accounts for the entropy 
of the initial black branes, the analysis between Eqs. (\ref{sugrent}) 
and (\ref{rcrit}) 
predicts that in this case the system will have the entropy-energy 
relation of a $\tilde{p}$-brane, up to a numerical factor of order 
unity.  
This can be verified explicitly by extremizing the entropy with respect to 
$N$ in our model, obtaining 
\be 
S\sim M^{\frac{c\left( p+1\right) 
     -\tilde{p}\left( c-1\right) }{\left( p-\tilde{p}\right) 
     \left( c-1\right) +p+1}}~. 
\ee 
In order for this to match  
the general formula (\ref{sugrent}) we must have that 
\be 
\frac{c\left( p+1\right) -\tilde{p}\left( 
c-1\right) }{\left( p-\tilde{p}\right) \left( c-1\right) +p+1}
  =\frac{D-\tilde{p}-2}{D-\tilde{p}-3}~, \ee
or equivalently,
\be 
\frac{p+1}{c-1}=D-3-p, 
\ee 
which is true 
\emph{precisely} for the values of $p$ and $c$ appropriate for the 
D3-, M2- and M5-brane cases.
 
 Now we have to check that the entropy really ceases to increase if 
 $L$ is of order $1/T$.  The brane-antibrane system has topology
 $\mathbf{T}^{\tilde{p}} \times \mathbf{I}^{p-\tilde{p}}$, where
 $\mathbf{I}$ is an interval of a certain length $L$.  
 Outside of this interval the tachyon is fully condensed and
 we have the closed string vacuum, so the 
 masses of all open string fields tend to infinity. 
 Because of this, the mass-terms in the action force all 
 fields to go to zero at the end points of $\mathbf{I}$. 
 This implies that there is no constant mode on $\mathbf{I}$, 
 and exciting any mode requires at least 
 an energy $1/L$.  If $T<1/L$ no modes are fully excited, and the 
 entropy decreases in a manner similar to what we saw
 in Section \ref{bhsec}, which is what we wanted to show.

 The possibility of describing lower-dimensional black branes in terms 
 of higher-dimensional ones is clearly an extension to the 
 non-extremal case of Sen's identification of D-branes as lumps in the 
 tachyon field \cite{sennonbps,dsol}.
 It is intriguing to note that in the case of eleven dimensions we 
 have found two different ways to obtain the lower-dimensional black branes.
 For instance, the black two-brane can either be viewed as pairs of M2's and
 $\overline{\mbox{M2}}$'s, or as pairs of M5's and $\overline{\mbox{M5}}$'s
 where the tachyon has condensed along three directions.

\section{Conclusions}
\label{5sec}

 In this paper we have studied brane-antibrane systems at finite 
 temperature.  It is intuitively obvious that a large enough 
 temperature will lead to the creation of \DD\ pairs, and as we have 
 seen in Section \ref{2sec}, the recent results in tachyon 
 condensation can be used to make this idea more precise.  The field 
 theory description allows in fact the creation of partially-condensed 
 brane-antibrane pairs, for which the expectation value of the tachyon 
 lies at neither the open string nor the closed string vacuum.
  As we have seen, in the canonical ensemble
  the system is unstable towards creation of an infinite number of 
  brane-antibrane pairs, so it should really be studied
  in the microcanonical ensemble, where only a finite 
  amount of energy is available.
  It is then favorable to put all of the energy into closed strings,
 rather than use it to create branes.  
 However, in the regime $g_s\ll 1$ states with $N\gg 1$
 \DD\ pairs and open strings can be long-lived, because the emission of 
 closed strings is suppressed.  If we keep $g_s N$ fixed, 
 then the open strings have non-trivial dynamics.
 We argued that the value of $\gs N$ and the energy density
 determine the nature of the 
 stable brane-antibrane state 
 (which is really meta-stable if we take into account the 
 closed strings).
 
 For $g_s N \ll 1$, there is a stable state with the tachyon
 partially condensed, that we described in Sections \ref{singlesec} and 
 \ref{multiplesec}.  This state contains partially annihilated \DD\ pairs 
 and massive open strings living 
 on them.  When 
 the temperature is of the order of $1/l_s$ (with a weak dependence on 
 $g_s$), the masses are of order $\sim 1/l_s$.  This fact might be 
 important when describing closed strings at high temperatures, as for 
 example in the investigation of the possible nature of the Hagedorn 
 transition, or near a black hole horizon.
 
For $g_s N \gg 1$, on the other hand,
we showed in Section \ref{stablesec} that the `tachyon' field is no longer
tachyonic at the open string vacuum
if the temperature satisfies $T>1/\sqrt{g_s N}$.
The \DD\ pairs then do not annihilate, and their  excitations
include massless open strings.  
The number $N$ of brane-antibrane 
pairs in this stable state 
cannot be chosen arbitrarily; it is determined 
thermodynamically by maximizing
the entropy keeping the total energy fixed.  
The entropy of the system is clearly zero if we do not create any 
brane-antibrane pairs (no degrees of freedom) or if we create the 
maximum number possible (largest number of degrees of freedom but no 
energy left to excite them), so the entropy will be maximized by an 
intermediate value of $N$ such that the mass of the branes and the energy 
of the open string gas are of the same order.  This state exists only 
in the non-perturbative regime $g_s N >1$, and it is therefore natural to 
associate it with a black brane solution of supergravity at the same 
temperature. 
Going back to the issue of closed string emission, in the supergravity
description that would correspond to Hawking radiation,
which is indeed suppressed for $g_s \ll 1$.

  In Section \ref{3sec} we examined more closely
  the identification between the 
  stable brane-antibrane state and a black brane in supergravity.
  Building upon \cite{hms}, 
  we formulated an explicit microscopic model 
  for the black three-brane in type IIB theory,
  and also for the two- and five-brane in M-theory.
  The model involved a stack of branes, a stack of antibranes, and 
  a gas  of massless open string modes on each stack.
  We argued that, at low energies, the theory on 
 the branes is decoupled from the theory on the antibranes.  
 We were then able to use the AdS/CFT correspondence to 
 compute the entropy of each strongly-coupled system 
 separately, and then add them together.  
 The resulting entropy agreed with the supergravity result 
 up to a puzzling factor of $2^\frac{p}{p+1}$, with $p$ the dimension of the 
 corresponding brane (in all cases this can be interpreted as a 
 factor of 2 in the energy).  
 Since the AdS/CFT correspondence uses 
 supergravity (in the near-extremal region), the result might appear 
 to be merely a  consistency check, 
 but the fact is that we are describing a very 
 different regime (including the Schwarzschild case),
 and a different situation, since the number of branes 
 can change.  The agreement we find is, at least to us, 
 a very unexpected property of the supergravity expressions.
 
 A puzzling fact is that, when the number of branes differs from 
 the number of antibranes, 
 to reproduce the supergravity result one has to assume 
 that the energy densities (or equivalently, the pressures)
 of the two gases are the same, implying that 
 their temperatures are different.  Since we argued that the theories 
 are decoupled, it is certainly not out of the question that the 
 temperatures could be different, but we lack an 
 understanding of why precisely the states with equal energies on the 
 two gases correspond to the supergravity solution.

  Since we obtain the correct entropy-energy relation, it is clear 
  that other properties of the black branes are reproduced.  In 
  particular, the specific heat of our microscopic system is negative. 
  When 
  the system loses energy, the energy density of the gas decreases, 
  which would seem to imply that its temperature decreases.  However, 
  the gas is no longer capable of sustaining the same number of branes.  
  A pair annihilates and the resulting energy is added to the gas, 
  increasing the temperature.  
  
  Another important property of the supergravity solution is the fact 
  that the black brane entropy increases as we decrease the volume.
  This is the thermodynamical formulation 
  of the Gregory-Laflamme instability.  
  If the black three-brane, for instance, 
  lives on a very large torus, then it has lower entropy 
  than a ten-dimensional black hole, so the latter is the
  preferred configuration.  In Section \ref{4sec}
  we showed that in our microscopic model it is also convenient to 
  reduce the size of the system, but only until it is of the order of 
  the inverse temperature.  Beyond that point the temperature is no 
  longer large
  enough to create excitations on the branes, so the entropy in fact 
  decreases.  The  
  radius that maximizes the entropy was found to 
  yield an entropy-energy 
  relation for the system 
  which is precisely that of a black hole in 
  ten dimensions, as expected from the supergravity side.  
  The entropy is proportional to $N^2$ and is 
  carried by the  
  components of the matrix fields, 
  rather than by their spatial dependence.  
  This is important since it means 
  that the gas on the branes does not probe distances smaller than the 
  Schwarzschild radius.  Contrary to what is expected for a normal gas, 
  increasing the total energy of our system
  does not help to probe shorter distances 
  (instead, the energy is used up in creating brane-antibrane pairs),
  in analogy with the behavior of a black hole.  
  
  The situation is different
  if the system lives on a sufficiently small torus. Since
  the branes can be multiply-wrapped, the large-volume 
  formulas are valid even if the temperature is smaller than the 
  inverse size of the torus. Brane-antibrane 
  annihilation would destroy the multiple wrapping and thus lower the 
  entropy, so the microscopic
  system is in this case stable, again in agreement 
  with the black brane in supergravity.

To summarize, the
field theory model we have studied in this paper
possesses several appealing features
which are in close correspondence with the properties
of black branes and black holes 
in supergravity.
Beyond the immediate task of questioning its assumptions and 
attempting to resolve the numerical discrepancy in the entropy,
there are many other aspects that remain quite obscure, 
in particular the usual questions about 
black hole formation, complementarity, etc. 
By analogy with the AdS/CFT correspondence,
it would also be interesting to see in what sense 
the near-horizon geometry of the black brane
(i.e., Rindler space)
is encoded in the microscopic description.  
Most important of all, 
the model should be tested in ways other than just 
computing the entropy.

For instance, similarly to \cite{msgrey} it will be interesting to compare 
the greybody factors with the field theory predictions.  
Also, one should be able to explicitly find 
the macroscopic perturbation of the system that 
is the equivalent of the 
Gregory-Laflamme unstable mode in supergravity.  
We believe that
these and other aspects of the model
merit further study.

\section{Acknowledgements}

We would like to thank Ansar Fayyazuddin, Joe Minahan, 
Subir Mukho\-padhyay, and Bo Sundborg
for valuable discussions.  
UD is a Royal Swedish Academy of Sciences
Research Fellow supported by a grant from the Knut and Alice Wallenberg
Foundation. The work of UD and AG was supported by the
Swedish Natural Science Research Council (NFR).

\appendix

\section*{Appendix A}

 In the text we found it
 useful to know all the components of the energy-momentum tensor rather
than just the energy. 
In order to do that we follow the procedure of \cite{dmrr},
where more details can be found.
We start by considering the D3-brane solution
\bea
ds^2 &=& h^{-\frac{1}{2}} ( - f dt^2 + dy_{\parallel}^2 ) 
  + h^{\frac{1}{2}} (f^{-1} dr^2 +r^2 d\Omega_5^2)~, \\
h &=& 1 + \sinh^2(\alpha)\frac{r_0^4}{r^4}~,  \\
f &=& 1- \frac{r_0^4}{r^4}~.
\eea
The solution can be written in isotropic coordinates 
by means of a change of radial variables
\bea
2\rho^4 &=& r^4 - \frac{1}{2}r_0^4 + \sqrt{r^8-r_0^4r^4}~, \\
ds^2 &=& h^{-\frac{1}{2}} ( - f dt^2 + dy_{\parallel}^2 ) 
+ \frac{r(\rho)^2}{\rho^2} (d\rho^2 +\rho^2 d\Omega_5^2)~.
\eea
For $\rho\rightarrow\infty$ we can write 
$g_{\mu\nu} = \eta_{\mu\nu} + h_{\mu\nu}$, with
\bea
h_{00} &\simeq& (1+\frac{1}{2}\sinh^2(\alpha)) \frac{r_0^4}{\rho^4}~, \\
h_{ij} &\simeq& -\frac{1}{2}\sinh^2(\alpha) \frac{r_0^4}{\rho^4} \delta_{ij}~, \\
h_{ab} &\simeq& \left(\frac{1}{4} + \frac{1}{2} \sinh^2(\alpha)\right) 
\frac{r_0^4}{\rho^4}\delta_{ab}~,
\eea
where $i$,$j$ are spatial indices parallel to the brane 
and $a$,$b$ denote the perpendicular directions.
 It is easy to see that $h_{\mu\nu}$ satisfies the harmonic gauge condition 
\be
\partial_{\lambda}h^\lambda_\mu - \frac{1}{2} \partial_\mu h = 0 ,
\ \ \ \ h = \eta^{\mu\nu}h_{\mu\nu}~,
\ee
which in turns simplifies the linear Einstein equations to 
\be
\partial_\lambda\partial^\lambda\left(h_{\mu\nu} 
- \frac{1}{2} \eta_{\mu\nu} h\right) = 16\pi G
T_{\mu\nu}~.
\ee
Inserting the asymptotic values of $h_{\mu\nu}$ we get
\bea
T_{00} &=& \frac{\pi^3}{\kappa^2} r_0^4 \left(\cosh(2\alpha) 
    + \frac{3}{2}\right) \delta(x_\perp)~, \\
T_{ij} &=& \left[\frac{\pi^3}{\kappa^2} r_0^4 \left(-\cosh(2\alpha) 
    + \frac{1}{2}\right)\right] \delta_{ij} \delta(x_\perp)~, \\
T_{ab} &=& 0~,
\eea
as used in the text. The delta-function 
in the transverse coordinates disappears after integrating
over a small volume
around the brane to get the $3+1$ dimensional $T_{\mu\nu}$.

 The same calculation can be done 
 for the M2- and M5-branes, to obtain the results used in the text.

\end{document}